\documentclass{article}

\usepackage[T1]{fontenc}
\usepackage[main=english]{babel}

\usepackage{times}

\usepackage{fancyvrb}
\CustomVerbatimEnvironment{CodeVerbatim}{Verbatim}{numbers=left, samepage=false}
\usepackage{float}

\usepackage{amsmath}
\usepackage{amscd}
\usepackage{color}
\usepackage{multicol}
\usepackage[margin=10pt,
        font=footnotesize,
        labelfont=bf,
        tableposition=top]{caption}
\usepackage{ifthen}
\usepackage{listings}
\usepackage{sidenotes}
\usepackage[utf8]{inputenc}

\usepackage{bookmark}
\usepackage[superscript]{cite}

\usepackage{xcolor}
\usepackage{xspace}
\usepackage{Sweave}

\DeclareRobustCommand{\DJ}{\marginnote{\large{\textcolor{magenta}{\sf Derek}}}}

\DeclareRobustCommand{\SC}{\marginnote{\large{\textcolor{blue}{\sf Stephen}}}}

\definecolor{hellgelb}{rgb}{1,1,0.9}
\definecolor{colKeys}{rgb}{0,0,1}
\definecolor{colIdentifier}{rgb}{0,0,0}
\definecolor{colComments}{rgb}{1,0,0}
\definecolor{colString}{rgb}{0,0.5,0}

{
}

\begin{document}

\title{A conversation around the analysis of the SiP effort estimation dataset}
\author{Derek M. Jones\\Knowledge Software\\derek@knosof.co.uk
\and Stephen Cullum\\SiP\\Stephen.Cullum@sipl.co.uk}
\maketitle

\begin{abstract}
The analysis of over ten years of commercial development using Agile
(10,100 unique task estimates made by 22 developers, under 20 project
codes) is documented via a conversation involving the data analyst
and a director of the company that created the SiP dataset.

Factors found to influence task implementation effort estimation
accuracy include the person making the estimate, the project
involved, and the propensity to use round numbers.

Any improvement in estimation accuracy, with practice, did not
noticeably improve regression models fitted.
\end{abstract}

\section{Introduction}

This paper takes the form of a conversation between Stephen Cullum
who started and ran the company that produced the SiP dataset
(software task implementation effort estimate/actual data) and Derek
Jones who analyzed the data.

Data analysis is an iterative process; ideas may have been suggested
by discussions with those involved before the data arrives, and new
ideas are suggested by feedback from the ongoing analysis.  Most
ideas go nowhere; failure of the data to support an idea is the norm.
Analysts who are not failing on a regular basis, never discover
anything.

The reason for doing data analysis is to obtain information that is
useful to those involved with the system that produced the data.

Any collection of measurements contains patterns, and some of these
may be detected the statistical techniques used.  Connecting patterns
found by data analysis, to processes operating in the world requires
understanding something about the environment and practices in which
the data was generated.

If the person doing the data analysis is not intimately familiar with
the environment and practices that generated the data, they either
have to limit themselves to generalities or work as a part of a team,
with people who have this knowledge.

The narratives created as explanations for the patterns found in the
data evolve as the conversation progresses; readers are not presented
with a well-structured story fitted together after the event.

The data is available at:
\url{https://github.com/Derek-Jones/SiP_dataset}.  The analysis was
done using {\sf R} and snippets of code appear throughout the text.

\subsection{Stumbling onto data}

\DJ

I first started talking to Stephen at a software engineering
workshop\footnote{The $56^\mathit{th}$ CREST Open Workshop,
\url{http://crest.cs.ucl.ac.uk/cow/56}}, and it sounded like the
company he had run for 16+ years had built up a very interesting
software project effort estimation dataset (as a side effect of
running the business).  It sounded to me that this data was much
larger and more detailed than anything currently available publicly,
and I told Stephen I would very interested in analyzing it.  Stephen
promised to investigate whether an anonymized form of the data could
be made available.

My professional background is compiler writing and source code
analysis.  Over the last seven years I have collected and analyzed
500+ software engineering datasets and made them publicly available
\cite{Jones_18}.

\SC

Software in Partnership Ltd (SiP) was started in late 2002 initially
providing back office insurance systems.  We were an early adopter of
agile methods and negotiated a green field development for a global
run-off and outsourcing provider, with the remit to centralise the
six separate systems they were using into one core product, providing
a licence cost saving alone of 1M per year on completion.  

One issue we had at that time, was that the quality of agile
management tools available were somewhat lacking.  The key proponents
of the agile process of the time actively recommending the use of
index cards and post it notes as the go to tooling (many still make
that same recommendation today).  We wanted something which was
flexible but also allowed us to learn from our mistakes.  We mimicked
the planning game in a custom-built application (Clarity) that
allowed each developer to pickup a Task, review the requirements and
provide an initial estimate to the client.  If approved the Task
locked the estimate and then allowed the developer(s) to record the
actual work carried out for eventual billing.  Once a Task was
completed, we had an Estimate vs. Actual for the developer(s) assigned
the work.  This approach allowed us to constantly review our
estimating capability (it was a part of our Friday 'how did it go?'
meetings).  The history of completed Tasks also helped in our
estimating process going forward, we would often be asked to do
something fairly similar to something we had done previously and
could use the actuals of those Tasks to guide our future planning.
Our Clarity application, provides the data discussed in this
conversation.

Prior to becoming IT Director at SiP I followed a pretty standard
career development path.  I had a fantastic first job working at a
dispatch company in the West End, carrying out pretty much any
technical work sent my way, from developing their in-house dispatch
system using a form of BASIC and machine code to drilling through
walls with a jack-hammer to wire up the network.  I subsequently
joined a large insurance systems provider and spent the next five
years of my career finding out how soulless software development can
be when done incorrectly.  I eventually jumped ship moving from the
provider to consumer side of the equation, ending up running the IT
function for a Lloyd's of London syndicate.  In my time here I
convinced the board to bring the majority of the outsourced IT back
in house and armed with a small team supported the business function
through software developed and managed directly by ourselves.

\subsection{The conversation}

\DJ

Getting the most out of data analysis requires domain knowledge.
Stephen has that knowledge, but is a very busy man.  I find the best
way to get a busy person to talk to me, is to tell them things they
find interesting and useful.

My top priority is to find something in the data that the domain
expert finds interesting.  The boring, but necessary, stuff can be
done later.

\subsection{The SiP dataset}

\DJ

The SiP dataset arrives and there is lots of it, i.e., 12,299 rows;
almost two orders of magnitude larger than the better estimation data
sets that are publicly available.

An initial analysis needs to answer two questions:

\begin{itemize}
\item do I believe the data?  This question is not about whether the
data was fabricated, but whether the information present is likely to
be a reasonably accurate representation of what it claims to be.
People make mistakes, decimal points may be misplaced, times and
dates entered later are misremembered, measurements are made using
different units by different people (e.g., miles vs.  kilometers),

\item what information is present in each column and what properties
does it have?
\end{itemize}

\begin{figure}
\begin{center}
\includegraphics{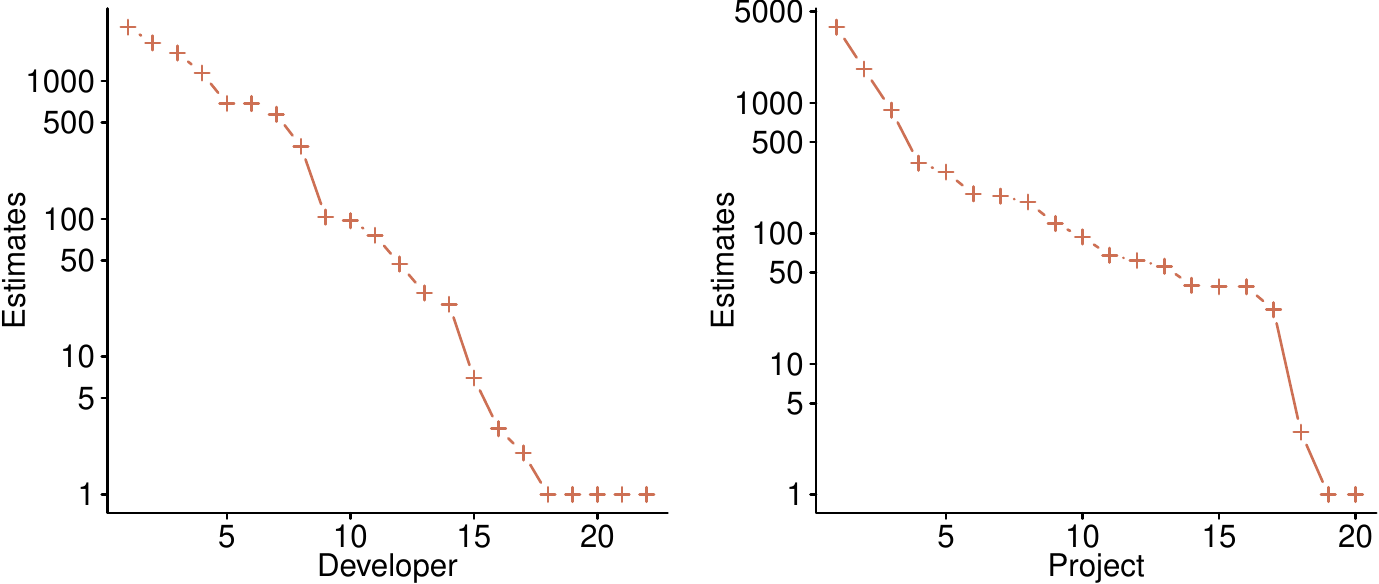}
\end{center}
\caption{Number of estimates by each developer and for each project in the SiP dataset.}
\label{Dev-Proj:fig}
\end{figure}

Between February 2004 and December 2014, 10,266 unique task estimates
were made by 22 developers, under 20 project codes (at the time of
the data snapshot, 1,848 tasks were still under development, and 166
had been cancelled).  Figure~\ref{Dev-Proj:fig} shows the number of
estimates made by each developer, along with the number of estimates
made for each project.

One method for getting a quick overview of data, is to look at it
within an editor (although, if there are lots of columns, line wrap
can complicate the visualization).  {\sf R}'s \texttt{str} function
lists column names, the base-type of the data and the first few
values.

\begin{lstlisting}[language=R,firstnumber=1,]
'data.frame':	12299 obs. of  20 variables:
 $ TaskNumber          : int  1735 1742 1971 2134 2251 2283 2400 2451 2475 2499 ...
 $ Summary             : chr  "Flag RI on SCM Message Summary screen using metadata RI application rules" "Allow RI Policies to be marked as Exhausted" "Fix Invalid UWREF Line DX402L99A1N" "New rows in the diary event for the SCM are read only." ...
 $ Priority            : int  1 1 2 5 10 1 5 5 6 5 ...
 $ RaisedByID          : int  58 58 7 50 46 13 13 13 13 1 ...
 $ AssignedToID        : int  58 42 58 42 13 13 13 58 13 26 ...
 $ AuthorisedByID      : int  6 6 6 6 6 58 6 6 6 58 ...
 $ StatusCode          : chr  "FINISHED" "FINISHED" "FINISHED" "FINISHED" ...
 $ ProjectCode         : chr  "PC2" "PC2" "PC2" "PC2" ...
 $ ProjectBreakdownCode: chr  "PBC42" "PBC21" "PBC75" "PBC42" ...
 $ Category            : chr  "Development" "Development" "Operational" "Development" ...
 $ SubCategory         : chr  "Enhancement" "Enhancement" "In House Support" "Bug" ...
 $ HoursEstimate       : num  14 7 0.7 0.7 3.5 7 7 7 1.4 1.75 ...
 $ HoursActual         : num  1.75 7 0.7 0.7 3.5 7 7 7 1.4 1.75 ...
 $ DeveloperID         : int  58 42 58 42 13 13 43 58 13 26 ...
 $ DeveloperHoursActual: num  1.75 7 0.7 0.7 3.5 7 7 7 1.4 1.75 ...
 $ TaskPerformance     : num  12.2 0 0 0 0 ...
 $ DeveloperPerformance: num  12.2 0 0 0 0 ...
 $ EstimateOn          : Date, format: "2004-02-26" "2004-02-26" ...
 $ StartedOn           : Date, format: "2004-02-26" "2004-02-26" ...
 $ CompletedOn         : Date, format: "2005-01-11" "2005-01-07" ...\end{lstlisting}

What do the columns represent?  The information present in some
columns can be guessed, from the column name (but it's always worth
checking for exceptions).

\SC
The columns extracted from Clarity provide the following information:

\begin{description}
\item[Summary] is a short text description of the action required to
complete the Task, specifically designed to be meaningful to the
client whenever possible.

\item[RaisedByID] the unique identifier of the individual who raised
the Task.  May be a client, or a member of SiP, depending on the
problem being addressed.

\item[AssignedToID] the SiP staff member responsible for the
completion of the work.

\item[AuthorisedID] the SiP manager with authority to sign-off a Task
on behalf of the client.

\item[StatusCode] the stage of completion the Task has currently
reached.  A Task has specific stages to progress through, as a new
stage is reached the business rules change, requiring further
information; with some fields becoming locked and others unlocking.

\item[ProjectCode] the specific work stream the Task is associated
with.  Many are client specific though there are a few, which are used
for the generic technical frameworks employed by SiP, or day to day
management such as holiday booking or operational management.

\item[ProjectBreakdownCode] a further breakdown of the work stream,
either to specific clients in a product with a varying client base or
sub systems in products built exclusively for one client.

\item[Category] an identifier categorising a management, operational
or development Task.

\item[SubCategory] a specific type of the parent category e.g.,
Management-Staff Recruitment or Development-Release.

\item[HoursEstimate] a decimal value representing the number of hours
estimated to carry out the work.

\item[HoursActual] a decimal value representing the total number of
hours it took to accomplish the Task across all SiP developers.

\item[DeveloperID] the unique identifier of the SiP developer who
carried out one or more actual items of work defined in the Task.

\item[DeveloperHoursActual] a decimal value representing the number
of hours a single SiP developer worked on the Task (possibly in
various roles).

\item[TaskPerformance] a decimal value representing the under-run (+
value) or overrun (- value) achieved by all SiP developers for a
finished Task.

\item[DeveloperPerformance] a decimal value representing the
under-run (+ value) or overrun (- value), for a single SiP developer
working on the finished Task.
\end{description}

\section{An initial analysis}

\begin{figure}
\begin{center}
\includegraphics{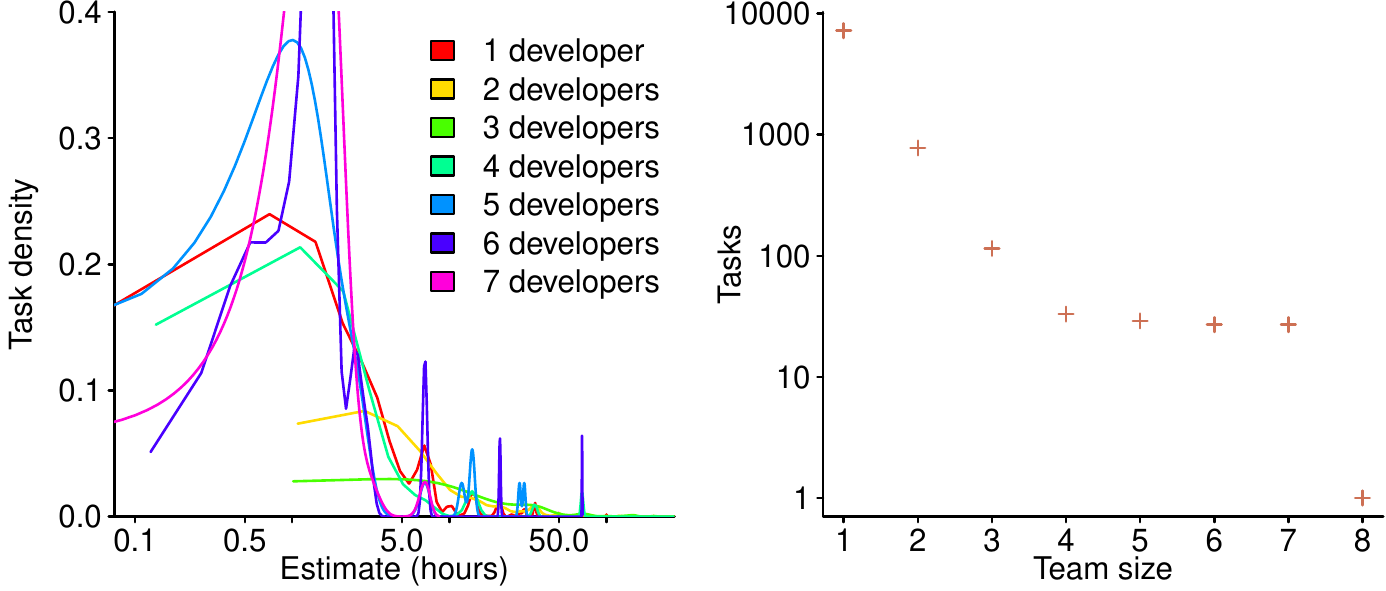}
\end{center}
\caption{Density of tasks having a given estimate duration, for teams containing a given number of developers (left); number of tasks involving a given number of developers (right).}
\label{team-est:fig}
\end{figure}

\subsection{A general snapshot}

\DJ

I was expecting tasks to involve multiple developers and take a day
or more to implement; incorrect assumptions on both counts.

Around 86\% of tasks involve a single developer, who makes the
estimate (see right plot in Figure~\ref{team-est:fig}).  For the
other 14\% of tasks the estimate is made by the team.

The left plot in Figure~\ref{team-est:fig} shows that task duration
ranges from under an hour to hundreds of hours.  I was surprised that
task duration appears to be independent of team size.  Note how the
density plots approximately mirror each other over a similar range of
estimates.

Task duration spanned two orders of magnitude (see left plot of
Figure~\ref{team-est:fig}).

\SC

SiP recorded any and all requests made by the user base, which were
reviewed after every full release.  Some requests could consist of
substantial amounts of work and would require a significant amount of
analysis to break the Task into a number of sub-Tasks.  These types
of request were recorded in a placeholder Task, which ensured the
idea was not forgotten and was assigned a rough estimate of the
effort we believed would be involved.  These placeholders were then
subject to periodic review, if the client stakeholders thought the
request worthwhile (perceived business benefit vs. estimated effort)
then time would be spent replacing the placeholder with a range of
more accurate sub Tasks (at which point the original Task would be
\texttt{CANCELLED}).  Occasionally, a Task with a rough high-level
estimate, would just get green lit.  In that the client agrees that
this work should be done and just wants SiP to get started, no
further analysis required.  This was a risk for us, just as much as
the client, but a long working relationship can foster this sort of
trust.

\DJ

The best way to understand data is to try to do something with it.

My experience is that jumping in and fitting a simple regression
model is a good place to start (my goto hammer for data analysis).

My reason for fitting a regression model is to gain understanding,
prediction is not of interest (at least not yet).

\subsection{The first regression model}
\label{first-regression-model}

\DJ

What is the relationship between \texttt{HoursEstimate} (estimated
hours, to implement the task) and \texttt{HoursActual} (actual hours
taken)?

One way of discovering relationships is to plot the data.

\begin{figure}
\begin{center}
\includegraphics{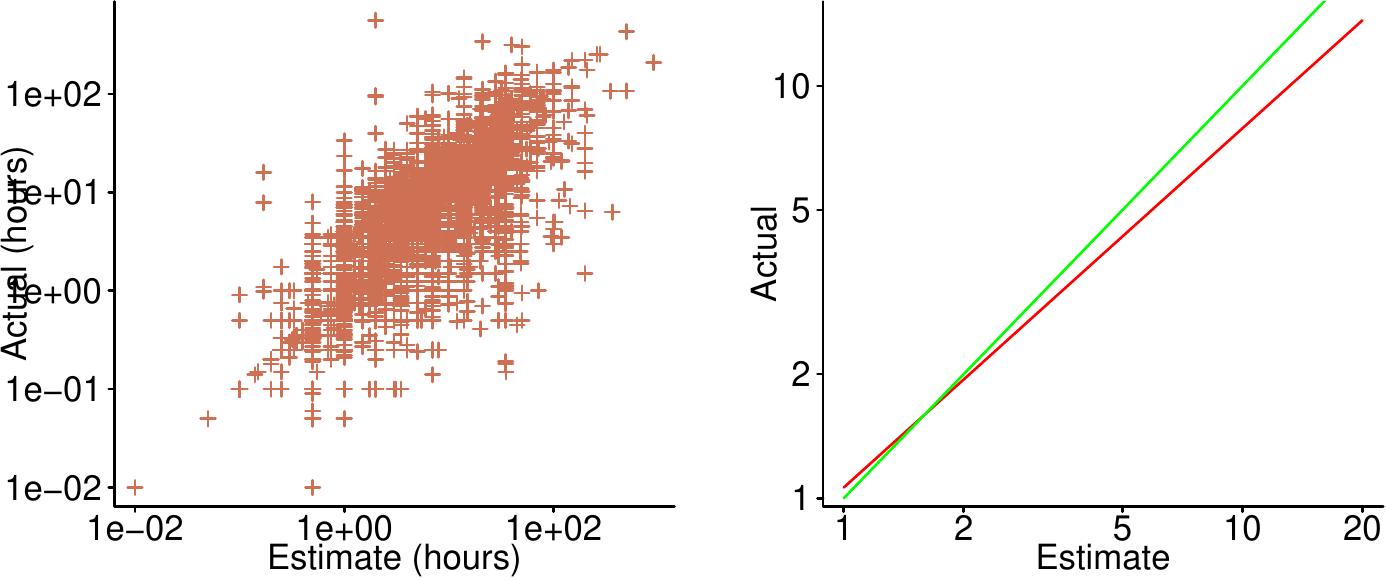}
\end{center}
\caption{Estimated against Actual effort, in hours (left); Line showing actual hours equals estimated hours (green) and fitted regression model (red).}
\label{Hours-Est:fig}
\end{figure}

The left plot in Figure~\ref{Hours-Est:fig} shows Estimated vs.
Actual effort (in hours) using log-scaled axes.

Fitting a regression model involving two variables is straight
forward:

\begin{lstlisting}[language=R,firstnumber=1,]
hours_mod=glm(log(HoursActual) ~ log(HoursEstimate), data=Sip_uTN)
\end{lstlisting}

This simple model explains over 70\% of the variance present in the
data; the coefficients of the relationship map to the following
equation:

$$\mathit{HoursActual}=1.1\times\mathit{HoursEstimate}^{0.87}$$

The numeric coefficients push in opposite directions: the $0.87$
exponent shrinks the value of $\mathit{HoursEstimate}$, suggesting
that developers are overestimating, while the multiplication by $1.1$
suggests they are underestimating.

The right plot in Figure~\ref{Hours-Est:fig} shows the ideal case
(green line) where actual equals estimated, and the fitted regression
model (red line).  The estimated hours are above actual, when
estimates are below 2-hours, but below actual when estimates are
above 2-hours.

Tasks estimated to take 1-hour, have an arithmetic mean for the
actual time of 1.36 hours (the geometric mean is 1.1 hours); for
tasks estimated to take 100-hours, the arithmetic mean of actual time
is 63-hours (the geometric mean is 38 hours).

What incentives are influencing the thinking process of the person
making an estimate?

It would not be cost effective to spend more time estimating than it
is likely to take doing the job.  Time spent estimating will be a
fraction of the likely estimation time.

A possible reason for short duration tasks to be underestimated is
that the person making the estimate does not spend enough time
studying the task to notice the potential pitfalls; intrinsic optimism
holds sway.  Stephen pointed out that short term interruptions are
unpredictable and can be a significant fraction of time on short
tasks.

The business context is the framework within which incentives
operate.  Is there an existing relationship between supplier and
client, and are estimates from multiple suppliers going to be
evaluated?

How changeable are clients requirements?  How reliable are suppliers
estimates?  An existing client/supplier relationship gives both
parties some idea about the answers to each others questions.  As
Stephen points out later in this conversation, SiP's ongoing
relationship with clients provides a foundation of confidence for
everybody involved.

When an estimate is part of a bidding process, there is a strong
incentive to produce a low estimate\cite{Flyvbjerg_13}; once a
project is underway the client has little alternative, but to pay
more (project overruns receive the media attention, and so this is
the more well-known case).

When an estimate is not part of a bidding process (e.g., internal
company projects, where the developer making the estimate may know
the work needs to be done), one strategy is to play safe and
overestimate, delivering under budget is often seen in a positive
light.  Underestimates receive little publicity, but are encountered
in studies of company internal tasks\cite{Hatton_07}.

\subsubsection{Using more variables to improve the quality of fit}

The \texttt{ProjectCode} column identifies the project associated
with the task.  It is possible that the accuracy of estimates will
vary between projects and this variable can be included in the model,
as follows:

\begin{lstlisting}[language=R,firstnumber=1,]
hours_proj_mod=glm(log(HoursActual) ~ log(HoursEstimate)+ProjectCode, data=Sip_uTN)
\end{lstlisting}

The fitted equation has the form (there is a small improvement in the
quality of the fitted model, as measured using AIC):

$$\mathit{HoursActual}=0.9\times\mathit{HoursEstimate}^{0.86}\times\mathit{ProjEffect}$$

where: $\mathit{ProjEffect}$ is a constant specific to each project;
its value varies from 0.6 to 1.5, for this data.

\begin{figure}
\begin{center}
\includegraphics[width=0.4\textwidth]{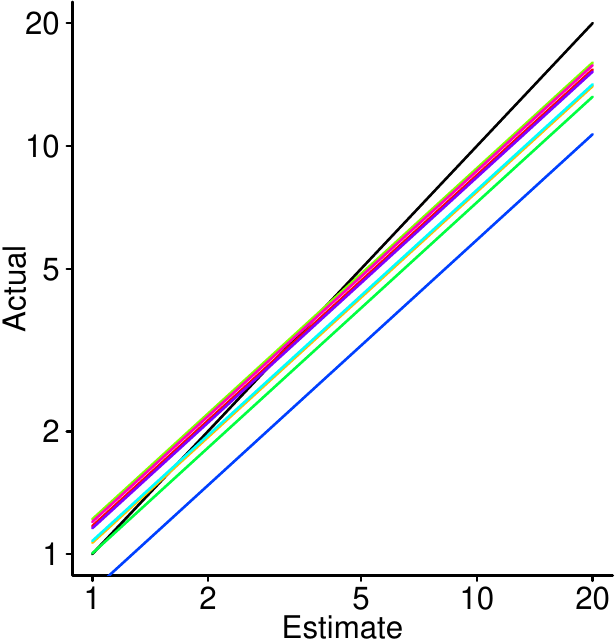}
\end{center}
\caption{Estimated against Actual effort, in hours, for tasks under \texttt{ProjectCode}.}
\label{Proj-Hours:fig}
\end{figure}

The lines in the left plot of Figure~\ref{Proj-Hours:fig} show the
equation fitted for each \texttt{ProjectCode} having at least 100
estimates; the black line is the case where Estimate equals Actual.
Points below the black line occur when Actual is greater than
Estimate, i.e., underestimates.  Projects vary in their cross-over
point, from under to overestimating.

There is an interaction effect between $\mathit{HoursEstimate}$ and
$\mathit{ProjEffect}$, and the much more complicated model taking
this into account changes Figure~\ref{Proj-Hours:fig}, such that the
lines were no longer parallel.

\SC

Each Task progressed through a number of stages, controlled by the
\texttt{StatusCode}.  The life cycle workflow followed the sequence:

\begin{lstlisting}[language=C,firstnumber=1,]
CREATED -> ESTIMATED -> AUTHORISE -> CHRONICLE -> COMPLETED -> RELEASED -> FINISHED
\end{lstlisting}

The flow could be interrupted at any point by \texttt{CANCELLED} but
any estimate or actuals allocated to the Task would remain for
analysis.  A Task would be \texttt{CREATED} to hold a description of
what was required and who raised it.  It would then be reviewed, and
the SiP employee responsible for doing the work would ensure the Task
was \texttt{ESTIMATED}.  The Task would then wait for appropriate
sign off.  Once \texttt{AUTHORISEd}, the Task would be in
\texttt{CHRONICLE} mode where actuals were recorded against the Task
by all SiP staff involved in the work.  Once the Task owner was happy
the change was implemented, the StatusCode was updated to
\texttt{COMPLETED}.  Some Tasks were effectively done at this stage,
as there was no artefact to release, however if something did need to
be pushed to the TEST server, or a document needed to be handed to the
client, then the StatusCode would be changed to \texttt{RELEASED}
once this was done.  All Tasks should have been set to the final
stage of \texttt{FINISHED}, once QA procedures were followed.
However, for many Tasks, \texttt{COMPLETED} was enough, and the
developers just left the Task at this status (which has proven
troublesome for the analysis within this paper, as discussed below).

\subsection{The practicalities of recording estimates}

\DJ

Would you expect all developers working on a Task to spend roughly
the same amount of time on it?

\SC

In theory yes if they were truly working together, but in practice
no.  Sometimes someone is simply being taken advantage of, with the
other developer letting them do all the work while they coast it.
Other times, one individual knows the sub system far better than the
other one and essentially teaches them in chunks.  One example task
is 3225 where the developer who knew about the warehouse did 75\% of
the work and handed the easier prep work to the junior.
Occasionally, they split the work rather than work together
(people!), and one individual comes in much later than the other
(breaking the estimate for both).  Sometimes someone is just helping
out when they have time, so will contribute less.  In the worst case,
the task is spiralling out of control, so they White Knight a
developer in, who is not included in the initial estimate but may end
up putting in much more effort than the task owner as they attempt to
unravel any problems caused by the first party.  

\DJ

How were individual developer estimates evaluated when they were part
of a team?  Was everybody held equally responsible for accuracy?

\SC

There had to be agreement that the estimate was achievable (before it
was reported to the client), but generally the most experience
developer set the pace.  However, if the estimate was not hit, then
the developer who owned the task was deemed to have failed (it was
their task after all).  This created quite an interesting dynamic as
junior developers being helped by senior developers could have some
spectacular overruns and feel they were led astray.  However, getting
estimates wrong in the beginning, when the tasks assigned were less
important, helped individuals new to the team focus on getting better
estimates the next time.  It also taught them to rely on their own
intuition and experience rather than relying on someone else.
Generally though, we had a good team dynamic and any ribbing was
constructive (all estimate fails [including mine] being reported at
the weekly meetings) and improving estimates was encouraged, rather
than being used to penalise people.

\DJ

Were estimate and actual hours rounded in any way?

\SC

The system did not round any actuals entered and despite Clarity
having a timer option, we generally didn't have a timer going (ala
Toggle).  Often despite working on your own work, as we were following
agile, you would get a shout out and if you could help, would pop
over to whomever was having trouble.  Normally, this was a 5-minute
interruption but sometimes it could be substantial.  We often forgot
to turn the timer on or off, so in the end didn't bother with it too
much as it didn't accurately record what we were up to.  The actuals
were normally entered at the end of the working day (you had to have
seven hours accounted for every day, otherwise all sorts of escalating
alert emails went off, starting at 5pm).  So, the numbers entered are
pretty accurate but subject to the same sort of human rounding that
the estimate value was.  

\section{Do people improve with practice?}

\DJ

In some activities, people get better with practice.  There is an
ongoing debate about whether the learning data is best fitted by a
power law or an exponential\cite{Heathcote_00}, but in practice there
is little difference in the fitted curves over the range of interest.

Does the accuracy of task estimates improve, as more are made?  Every
task has an associated number, \texttt{TN}, which increases over
time, and individual developers may improve with practice; a model
can be fitted for each developer.

Prior developer estimation experience is unknown, but is assumed to
be non-zero.  For the purposes of model fitting an initial experience
of 100 is assumed, for estimating practice (changing this value does
not have much impact), and every estimate treated as an opportunity
to practice.  An exponential rate of learning is assumed, so skill is
expected to improve as the $\log$ of the number of learning
opportunities.  The following code fits a distinct model for each
developer:

\begin{lstlisting}[language=R,firstnumber=1,]
df$TN=100+(1:nrow(df))

ea_mod=glm(log(HoursActual) ~ log(HoursEstimate)+ProjectCode+log(TN), data=df)
\end{lstlisting}

The quality of each fitted model is better (measured by the variance
explained) than the model fitted to all the data.  This is to be
expected since the models are tailored to the individuals making the
estimates.

The fitted model coefficient for \texttt{log(TN)} varies between 0.4
and 2.5.  This implies that with experience, some developers
increasingly overestimate and some increasingly underestimate.  This
does not sound right, perhaps the model is too simplistic and is
failing to take important factors into account.

\SC

For clarification there were no direct financial penalties incurred
for the company if our developers overestimated/underestimated the
effort required for a task.  There were plenty of political/cultural
ones though (along with financial side effects), such as being viewed
as slow or incompetent.  Excessive overestimates ran the risk of the
client simply mothballing that part of the development or seriously
curtailing the functionality, limiting the effectiveness of the
product overall (and affecting income for that part of the system).
However, the effect of underestimates were much more troublesome, in
an agile environment where the development team sit with the users of
the software, there is nowhere to hide.  If you agree to deliver some
functionality within a week and that overruns significantly then you
need to prepare for some pretty serious conversations.  As this was
our first agile development with paying clients (some of us had been
involved in agile projects for the companies we previously worked
for; but in this instance you are immediately viewed as being 'on the
same side'.  This is certainly not the case in the early stages of an
agile commercial project.) the shock of developers who have
accidentally underestimated (with the best of intentions), suddenly
being faced with an angry client (who is not directing their anger
through a hierarchy of project and business managers from previous
development models but has simply rocked up at their desk) is a sight
to behold.

As the development team became aware of their high visibility, there
was a change in how we approached estimating.  Initially (and I
suspect this is true of every development team) we wanted our clients
to like us and would provide some close to the wire estimates as it
was the easiest way to keep them happy.  We eventually realised this
was a false economy, either we would end up carrying out death
marches to deliver the code, becoming fatigued in the process and
having to explain to the clients finance department why the hours
worked that month were far in excess of what they were expecting (and
were willing to pay for) or we would fail to deliver, perhaps having
to field complaints from the client, but more importantly watch the
user(s) we disappointed have less faith in our ability to deliver and
as a consequence engage less with us, and the product going forward.
We became significantly more risk adverse.  We expected unexpected
issues to occur frequently and as a consequence would stand firm
against client pressures to squeeze more into a release.  Our clients
became very unhappy during the planning stages of a release cycle, as
they felt we were not really listening to what they needed to do
their job.  We often had a number of sessions trying to gain
consensus on the bare minimum we could deliver that would provide
business benefit.  However, (and we were not really expecting this),
they became very confident in our abilities to manage and run the
project when release after release we delivered what we said we
would.  When we occasionally had spare capacity in the cycle and
included unexpected functionality for the release [which we nicknamed
'developers choice' (as the developers with the free time got to
choose)] the clients were genuinely very pleased, with that goodwill
feeding back into the team. 

Our approach simply moved client dissatisfaction from the end of a
release cycle to the beginning.  Hence, our approach became to under
promise and over deliver.  That is not to say we didn't make some
serious underestimating howlers, we did (and still do), but it became
clearly preferential to be honest with the users upfront (potentially
disappointing them then), to providing false hope and disappointing
them later.  By under promise I mean we generally added a tolerance
of 5-20\% (dependent on the complexity of the task) rather than
simply doubling what we originally thought of.  With respect to
Derek's findings above I would concur that experienced developers
exposed directly to and working with a reasonable user base would
tend to overestimate (to a degree) based on a clients reactions to
late delivery. However, I am talking a considered overestimate to
ensure the majority of tasks are delivered on time (or very close to
it - no death marches).  Bearing in mind the fallout of continual
late delivery the instances of developers increasingly
underestimating does seem anomalous.

\DJ

\begin{figure}
\begin{center}
\includegraphics{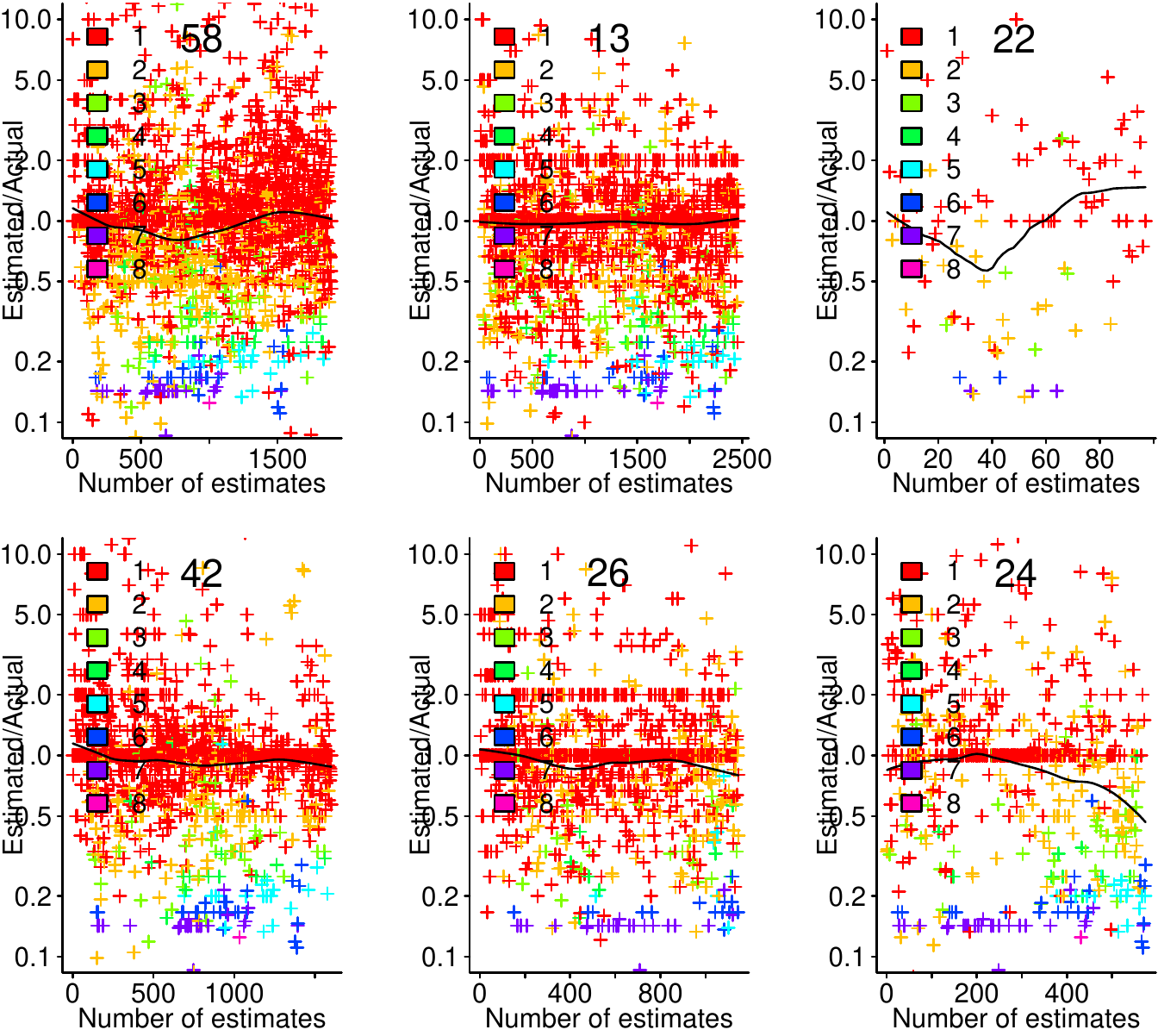}
\end{center}
\caption{Accuracy of estimates made by some developers.}
\label{dev-est:fig}
\end{figure}

How does the accuracy of developer estimates vary over time?  Each
plot in Figure~\ref{dev-est:fig} shows the ratio estimated/actual for
successive estimates made by one developer (selected because they had
made the most estimates).

One possible explanation for the fluctuating, small changes in
performance is that as developer experience increases, the time taken
to implement tasks decreases.  If increased developer experience
causes them to give lower estimates, then estimation accuracy may
remain unchanged.

\subsection{How SiP tracked work}

\SC

In our day to day work clients and staff worked at the Task level.
By that I mean if a Task came in 30 minutes early, they didn't assume
that those 30 minutes, would be used on another Task automatically,
perhaps the developer would spend time chatting with a user, carry
out housekeeping, return phone calls or emails or any other
non-chargeable Task you can think of. If it came in late, then the
additional time was generally found by pushing back work outside of
the release or taking the work home with you.

If the Tasks agreed for the release were completed by the deadline,
then everyone was happy.  This meant that under estimated work was
bad but accurate and over estimates were deemed OK, i.e., we hit the
target or came in earlier (and cheaper).  We reported Tasks to the
client not hours because that is what they cared about.  This meant
we thought of Tasks in terms of accurate, under and over estimates,
rather than hours, minutes and seconds.  

Carrying out a very crude timeline analysis of \texttt{COMPLETED} and
\texttt{FINISHED} Development Tasks across all SiP staff shows some
interesting behaviour.

\begin{figure}
\begin{center}
\includegraphics{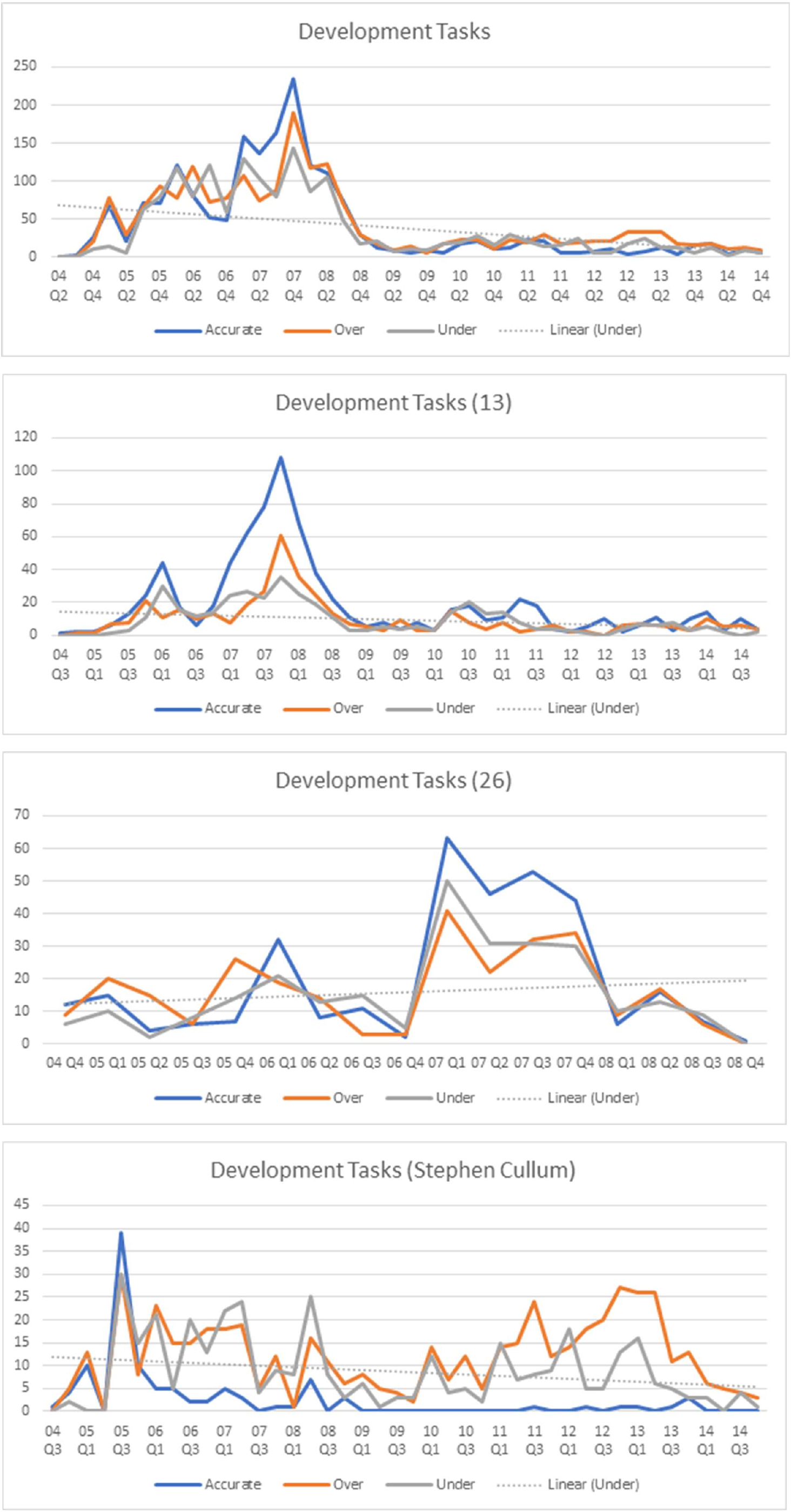}
\end{center}
\caption{Number of accurate, over and under estimates, for each quarter; top all developers, others are specific \texttt{DeveloperID}s.}
\label{cou-stephen:fig}
\end{figure}

In the heaviest development times (first 5-years) over and under
estimates flip flopped, which lead to improved accuracy.  After the
first 5-years when the majority of development tailed off into the
support phase, work volume dropped but interestingly so did the focus
on accuracy, with over estimating taking precedence.  I am assuming
we became more aware of the political benefits of under promising,
whilst over delivering.

Taking two active developers, say 13 and 26 this pattern repeats
(see Figure~\ref{cou-stephen:fig}).

Accuracy improves over time, especially in heavy development phases
(2005-2008), but as supporting the product becomes the key activity,
accuracy loses its prominence.  Generally, accuracy and over
estimating prevail but occasionally the developer has a phase of
under estimating but that does seem to get corrected over time.

Interestingly, my behaviour is quite different to other staff, see
Figure~\ref{cou-stephen:fig}.

Very early on I seem to have decided an over estimate was preferable
to an accurate one! (As well as the client good will, it provides a
buffer to any of my under estimates.)

Very roughly, about 30\% of our estimates are under estimates, which
is consistent across all staff.

I believe there was learning going on, but not necessarily to improve
estimates by reducing the risk buffer.  We learnt that whilst
accurate estimates are king in hard core development phases, over
estimates are beneficial in support phases.  Over estimates where you
simply procrastinate to pad the time is unethical behaviour and
simply becomes an 'accurate' estimate that is costly to the client.
An over estimate whereby, if you avoid any unanticipated delays, you
simply pass that gain straight back to the client (in terms of
reduced cost, fresh development staff or additional Task pickup for
that release) is a way to accept a margin for risk reduction but
reward the client if the risk does not materialise.

\section{Coworker influences}

\DJ

The previously fitted model treated developers as individuals, not as
possible members of a team.

Teams are small (78\% of multi-person teams contain two people),
which means there is likely to be an impact due to individuals and
perhaps a team size effect.

The following code adds team size (i.e., number of developers); some
trial-and-error experimentation (i.e., 0.5 and -1.0). was used to
select the exponent for team size.  Adding a variable for the
cumulative number of estimates made by developers on the team has a
minuscule impact on the fitted model:

\begin{lstlisting}[language=R,firstnumber=1,]
ea_form=formula(paste("log(HoursActual) ~ log(HoursEstimate)+",
					ProjectCode+I(size^0.5)+",
					paste0(udevID_str, collapse="+")))

ea_mod=glm(ea_form, data=unq_sip_exp)
\end{lstlisting}

The fitted equation, which explains 77\% of the variance in the data,
is:

$$\mathit{HoursActual}=0.15\times\mathit{HoursEstimate}^{0.82}\times\mathit{ProjEffect}\times\mathit{DevEffect}\times e^{6.7\sqrt{\mathit{TS}}}$$

where: $\mathit{TS}$ is a task's team size, and $\mathit{DevEffect}$
is the combined influence of all developers on the team.

Points of interest include:

\begin{itemize}
\item

adding the square-root of team size to the model improves the quality
of fit by 1\% (this aspect of the model is dominated by the 86\% of
teams consisting of a single developer),

\item estimates from some individuals were consistently much
lower/higher than the average team estimate,
\end{itemize}

Derek sent Stephen a list of \texttt{DeveloperID}s and the model's
prediction concerning the lower/higher than SiP developer average
characteristic of their estimates.

\SC

The SiP Clarity system allowed staff to collaborate on a given tasks,
so actuals could be recorded for each category of effort carried out
by each individual working on the Task.  For example, individual 1
could be enhancing the code (adding features), individual 2 could be
updating the application documentation, whilst individual 3 could be
responsible for manually testing the changes.  A Task could hold a
range of actuals, which were representative of an individual's
involvement.  We could also cater for Pair Programming but there was
not an explicit category for that (both developers would enter the
same category Enhancement or Bug).

The system did not have this flexibility for recording the estimate.
A multi estimate mechanism, whereby everyone involved could provide
their own estimate (for their involvement), before the total was
supplied to the client would have been a better instrument for
improving estimating skills across the team.

I decoded the serial over and under estimators (based on their
\texttt{DeveloperIDs}).  Derek identified one software developer who
consistently overestimated.  This uncovered nothing exceptional other
than a junior developer (first software development position) who was
ultra cautious.  Derek also identified two SiP staff members who were
serial under-estimators, their business function was client
management/sales and marketing.  The two roles were not subjected to
weekly peer review (they were not a chargeable resource).  SiP
required all staff to use Clarity, but the focus on Estimates vs.
Actuals was a function of the software delivery process and internal
tasks were not subject to this review.  This is the probable
explanation as to why these individuals were so poor at estimating,
they were never challenged on their estimates so there was no need to
learn from the artefacts they had previously created (although
self-improvement would suggest you should try).

\section{Serial correlation of estimates}

\DJ

When estimating, are developers influenced by recently completed
estimates?

Possible causes of serial correlation, in task estimation times,
include:

\begin{itemize}
\item Anchoring\cite{Lieder_18} is the term used to describe the
situation where a statement given to a person prior to asking them a
question, influences the answer they give.  For instance, asking:
\textquotedblleft{}Did you on average write more or less than $N$ Lines
of Code per work-hour in your last project\textquotedblright{}, where
the value of $N$ varied between 1 and 200, changed the mean response
by 72 lines\cite{Jorgensen_12},

\item work on related tasks may be ordered to fill a complete day or week,
or to fill in time remaining in a day or week,

\item the estimates are often integers and six values account for 54\%
of all estimates.  The correlation may be an artefact of the
probability of identical values appearing in sequence.
\end{itemize}

The following list shows the 10 most frequent pairs and triples of
the same sequence of estimates (\texttt{prop} is the proportion).

\begin{verbatim}
   ngrams freq   prop     ngrams freq    prop
1    1 1   256 0.0249     7 7 7    55 0.00536
2    7 7   199 0.0194     1 1 1    45 0.00438
3    1 2   176 0.0171     1 2 2    40 0.00390
4    2 1   170 0.0166     7 7 1    34 0.00331
5  0.5 1   165 0.0161   0.5 1 1    34 0.00331
6  1 0.5   164 0.0160     1 2 1    32 0.00312
7    2 2   134 0.0131     2 2 1    29 0.00283
8    7 1   131 0.0128   1 1 0.5    27 0.00263
9    1 7   118 0.0115 0.5 0.5 1    26 0.00253
10   1 3   114 0.0111     2 1 1    26 0.00253\end{verbatim}

Given 1,525 estimates of 1-hour and 8,741 non-1-hour estimates, how
many 1-hour estimate pairs are likely to occur?  The number of pairs
is 1,525+8,741, and the probability of a pair occurring is:
$\frac{1525}{1525+8741}\times\frac{1525-1}{1525-1+8741}\to 0.022$,
giving $(1525+8741)\times0.022 \to 226$ pairs; 12\% less than the
pairs of 1-hour estimates that occur.

Table~\ref{Est-pairs:tab} shows the expected number of pairs of the
same value, for the ten most common estimate values.  The number of
pairs calculated is much lower than those appearing in the data.

If the correlation seen is a consequence of the sample containing a
few frequently occurring values, the correlation will be present when
the values are randomised.  Randomising the sample, calculating the
correlation, and repeating the process many times, finds that the
mean correlation is almost zero (with a standard deviation close to
zero).

The correlation present is not the result of the series containing a
few frequent values.

\begin{table}[ht]
\centering
\begin{tabular}{rrrrr}
  \hline
1 & 2 & 7 & 0.5 & 3 \\ 
  \hline
226.41 & 99.67 & 96.74 & 84.35 & 41.73 \\ 
   \hline
\end{tabular}
\caption{Expected number of pairs of the same estimate value.} 
\label{Est-pairs:tab}
\end{table}
I have found some serial correlation, i.e., the estimate for the
previous task sometimes seems to have had an impact on the next
estimate.

\SC

This is interesting, I was not expecting that for the Estimates.  I
would have expected it for Actuals as we tended to group specific
sorts of tasks together and process them one after the other.  For
example, we often got various requests made throughout the week for
changes to be made to metadata (say fields available in our Search
Control) each of these Tasks were quick to implement (< hour) so we
saved them up (generally for an easy win on Fridays).  We had much
less control over the estimates though, as we did not know when the
clients would turn up at our desks and request something.  Whoever
was approached had the job of raising the task, prior to the estimate
being added at the next planning session.  Perhaps at the planning
sessions we grouped alike tasks as well, but subconsciously as it
certainly wasn't something we did intentionally.

\DJ

I later found a mistake in my analysis.  In the data, tasks involving
multiple developers appear multiple times, one row per developer.
Filtering, so that each task is represented by one row, halves the
serial correlation.  The signal in the data suggesting a possible
$AR(1)$ model, disappears.

\section{Round numbers}

\DJ

When giving a numeric answer to a question, people sometimes choose
to give a value that is close to, but less exact, than what they know
to be true; values may be rounded towards a preferred value, known as
round-numbers.  Round-numbers are often powers of ten, divisible by
two or five, and other pragmatic factors\cite{Jansen_01}; they can
act as goals\cite{Pope_11} and as clustering
points\cite{Sonnemans_06}.

\begin{figure}
\begin{center}
\includegraphics{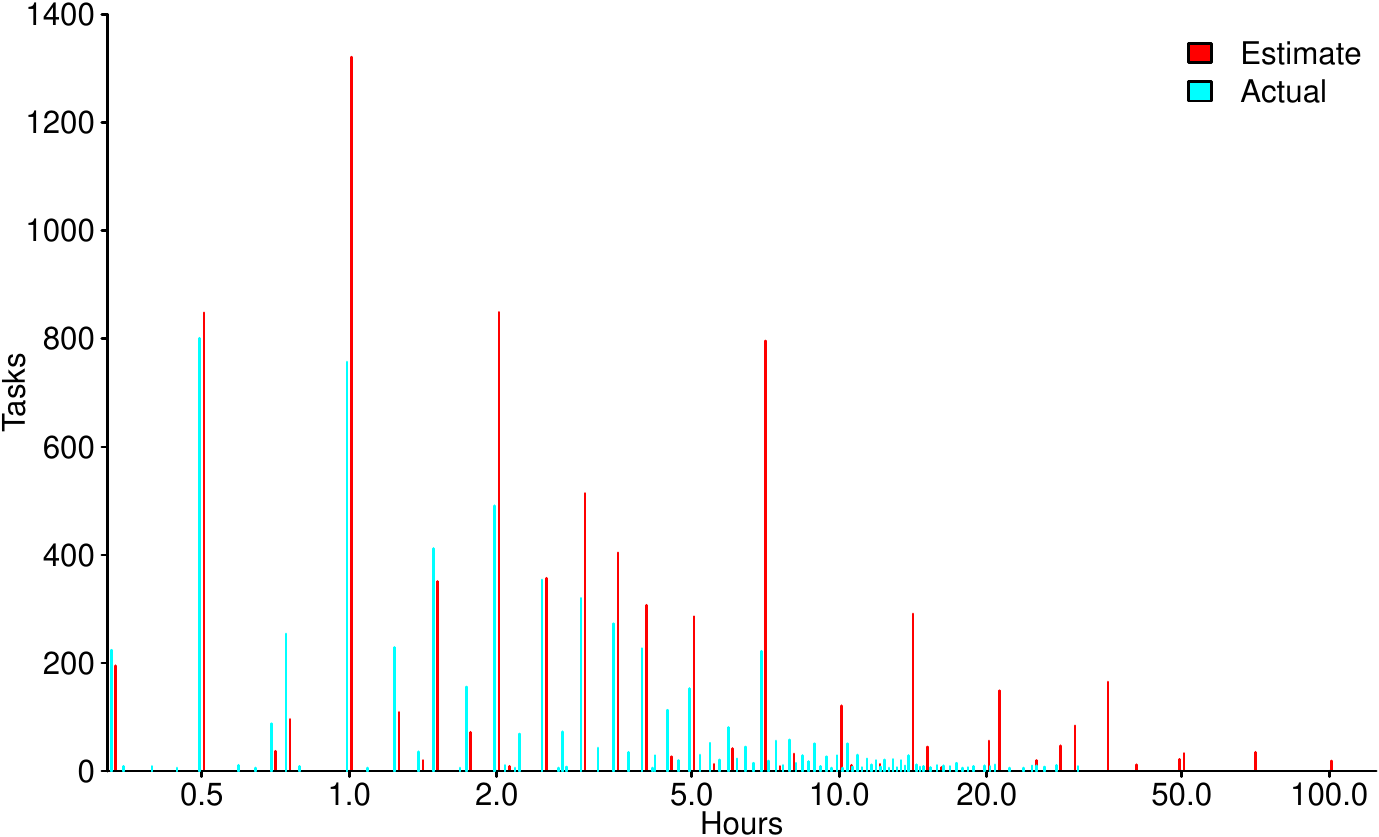}
\end{center}
\caption{Number of Estimates (red) and Actuals (blue) at a given number of hours (slightly offset to prevent overlaps obscuring information).}
\label{Round-Est:fig}
\end{figure}

Figure~\ref{Round-Est:fig} shows the total number of estimates (red)
and actuals (blue) having a given value (lines are slightly offset so
information is not obscured).

There are estimate peaks at 7, 14, 21 and 28 hours, suggesting that
some developers estimated in days and then converted to hours.  There
are 35 hours in a week, and the 300+ estimates for this value suggest
that one week is a round-number for some task estimates.  The
estimate peaks at the round values 10, 15, 20 and 30 hours are
smaller than the 'day' peaks, but still larger than adjacent
non-round values.

Regression models built using hour round-numbers and 'day' hours
(having the same range of hour values as the hour round-numbers) are
very similar (see below), i.e., if developers are thinking in units
of hours and days, the estimates are not significantly affected.

$$\mathit{HoursActual} = \begin{cases}1.27\times\mathit{HoursEstimate}^{0.80}&\text{round-number hours}\\
		1.31\times\mathit{HoursEstimate}^{0.77}&\text{'day' hours}\end{cases}$$

\subsection{The impact on accuracy}

\DJ

It looks like low hour estimates are balanced under/over estimates,
but the larger estimates are biased towards overestimating.

\SC

This makes sense as the low estimate work was generally something we
were comfortable with and probably did fairly frequently (essentially
plug \& play code into our framework).  Larger estimates were normally
a big block of functionality that did not exist in our framework
already and was therefore more technically complex with unknowns.

\DJ

The 1-hour estimates would have been more accurate if they had been
1.4 hours (the arithmetic mean of the actuals; the median is 1-hour,
i.e., there are as many overestimates as underestimates); the
corresponding value for the 2-hour estimates is 3.3 hours.  Let's say
you were given this analysis after running the company for a year,
i.e., based on a year's data.  What would you have done with it?

\SC

This is a difficult one to answer as the client, but moreover the
users of the system, were fundamentally happy with how we proposed
and delivered work.  We released to production every 3-4 weeks, so it
was unusual for a slippage on a single Task to affect them (but not
unheard of).  Generally, where we lost on one Task, we gained that
back on another or if the release deadline was approaching, and a Task
looked shaky we could increase developer effort (which could be
cashed in after the release).  In cases where we failed to deliver on
a key specific Task for that release (i.e., it was negatively
affecting the business), we could always push out an emergency
release.  But essentially, system users were used to waiting for a
Task, so if we overran an hour estimate by 100\% or delivered in half
the time, the user who raised the Task would not see it until the
next release anyway, so we had a recovery buffer, which withered as
we got closer to the release deadline.

Estimate quality affected the SiP Developers ability to decide how
much we squeeze into a release, much more than the users of the
system.  If we got it wrong, then as professionals we suffered the
consequences, rather than passing them onto our clients.  By that, I
mean, if my estimates were off, I would often work overtime to ensure
that the Task was still delivered to the client for the release (as
would the majority of our developers).  

Increasing estimates from 1 to 1.4 hours (or 1.5 to keep to client
hours of 1 hour 30 mins) would have had minimal impact.  Going much
higher, I think would have raised concerns, especially as, after a
while, the users become familiar with how we estimated the simpler
work.  Given the information I would have fed back to the developers
that unless they were really confident (i.e., lots of evidence in
Clarity [where they are doing the work]) then low end Tasks would
benefit by adding an extra half an hour to them.  If the client
queries any specific Task for the increase, then we can provide
evidence as to why it is more accurate than what we were previously
providing.  So yes, I would have taken action on the information but
would have monitored client reaction at the planning meetings quite
closely.  What we see as an informed decision could be seen as a way
to squeeze additional money from the client, so handling this sort of
change delicately is paramount.

\DJ

Picking round numbers, rather than more accurate estimates, has the
advantage that clients may find round numbers more believable. Client
influenced estimates are a perennial problem.  Getting developers out
of a round number mindset is not useful if the client thinks in round
numbers.  There are very few estimates for 6-hours, but lots for 5
and 7 (1 day).  This is obviously a rounding effect, which decreases
estimate accuracy, but it might make clients happier (or less
unhappy), whereby little effort is likely to be investigated in tasks
thought to require a short amount of time.  More time will be
invested to estimate tasks likely to require many hours.

\SC

Nice insight Derek, we often think of how we affect business as they
interface with us, but it is much subtler the other way around.  Here
is a clear indication that we became naturalised to the client's ways
of working/thinking despite working within an Agile framework
(something that was alien to the client).  I had never considered
this before.

\section{More data becomes available}

\begin{figure}
\begin{center}
\includegraphics{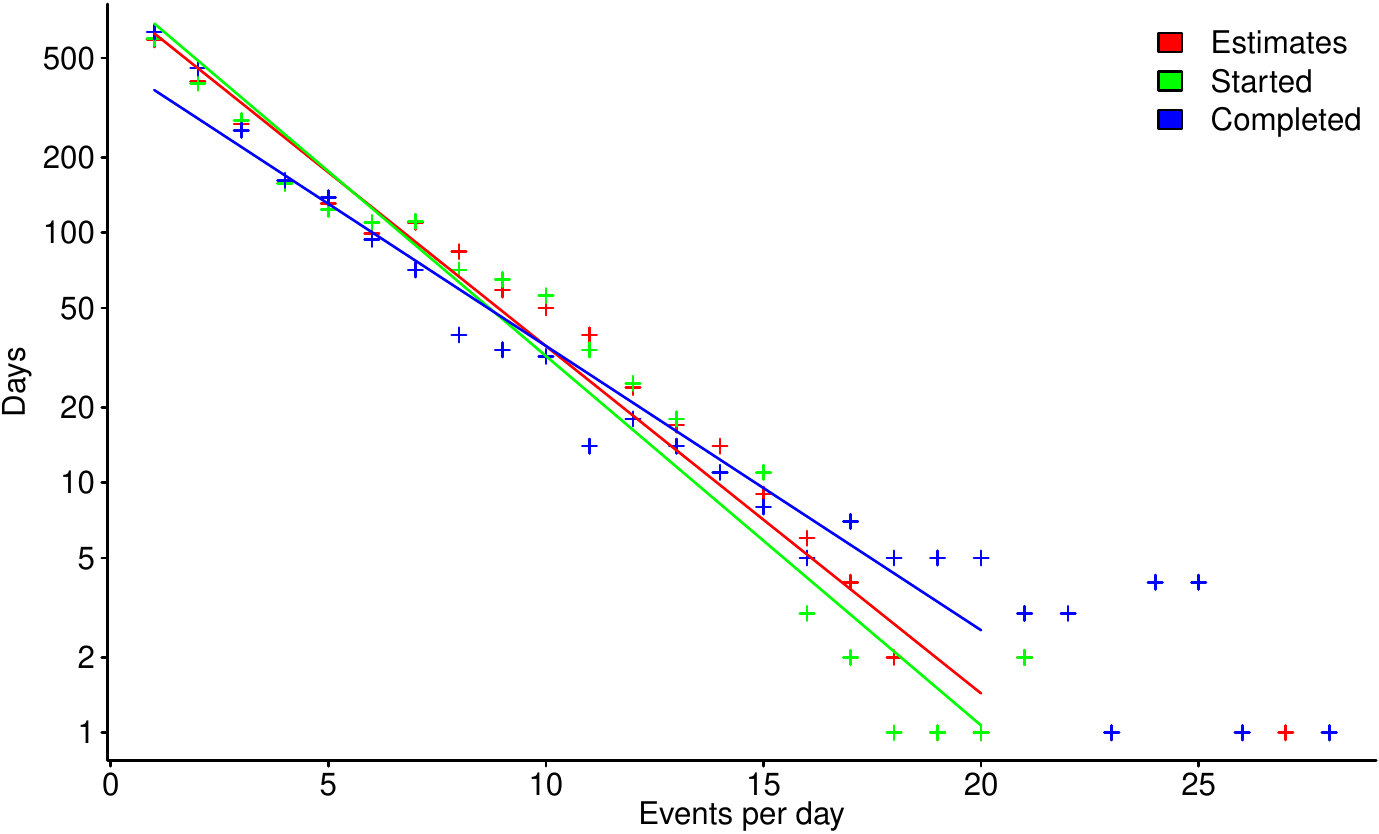}
\end{center}
\caption{Number of days on which a given number of task estimates, starts and completions occurred, with fitted regression lines.}
\label{Events-per:fig}
\end{figure}

\DJ

Feed your supplier of data interesting information, and they are more
likely to be willing to spend time tracking down more data.

The new data contains three columns: the date on which the Estimate
was made, work started on the task and when the task was completed.

Figure~\ref{Events-per:fig} shows the number of days experiencing a
given number of events (task estimate, start, completion), with
fitted regression lines.  The numbers for tasks estimated and started
are very similar because for 90\% of tasks, both events occur on the
same day.

\section{Time distributions}

\subsection{Distribution of actuals}

\DJ

How are the actual task hours distributed about the estimate for a
task?

\begin{figure}
\begin{center}
\includegraphics{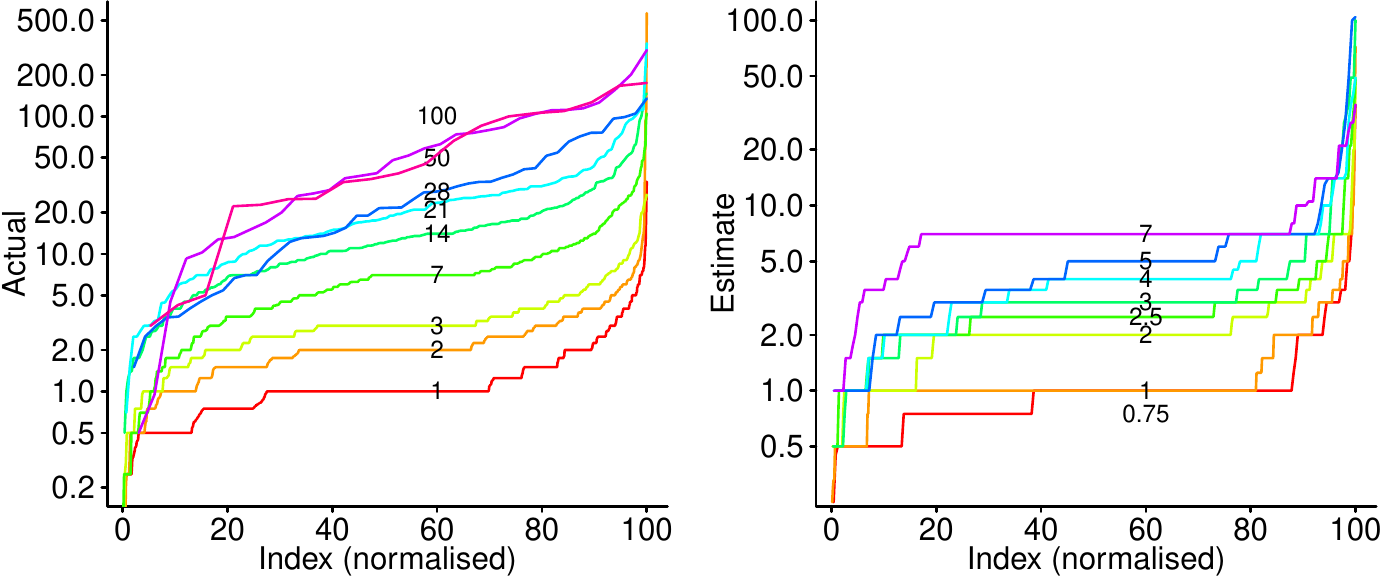}
\end{center}
\caption{Actual hours for specific estimate values (number on line, vertically above 60), x-axis normalised (left); Estimated hours for specific actual values (number on line, vertically above 60; right).}
\label{Act-per:fig}
\end{figure}

The left plot in Figure~\ref{Act-per:fig} shows actual task hours for
common estimate values (given on each line, vertically above the 60).

The distribution of actual hours, for a given estimate value, changes
with the magnitude of the estimate.  It is possible to create a list
of task characteristics that change with task duration, but without
detailed information about the application and task implementation,
explanations are essentially just-so stories.

\subsection{Distribution of estimates}

\DJ

How are the estimated task hours distributed about the actual hours
for a task?

The right plot in Figure~\ref{Act-per:fig} shows estimated task hours
for common actual values (given on each line, vertically above the
60).

\subsection{Distribution of elapsed working days}

\DJ

While 51\% of tasks were started and completed on the same day, some
relatively low effort tasks experienced much longer elapsed times
(i.e., multiple working days); 89\% of tasks were estimated and
started on the same day.

\begin{figure}
\begin{center}
\includegraphics{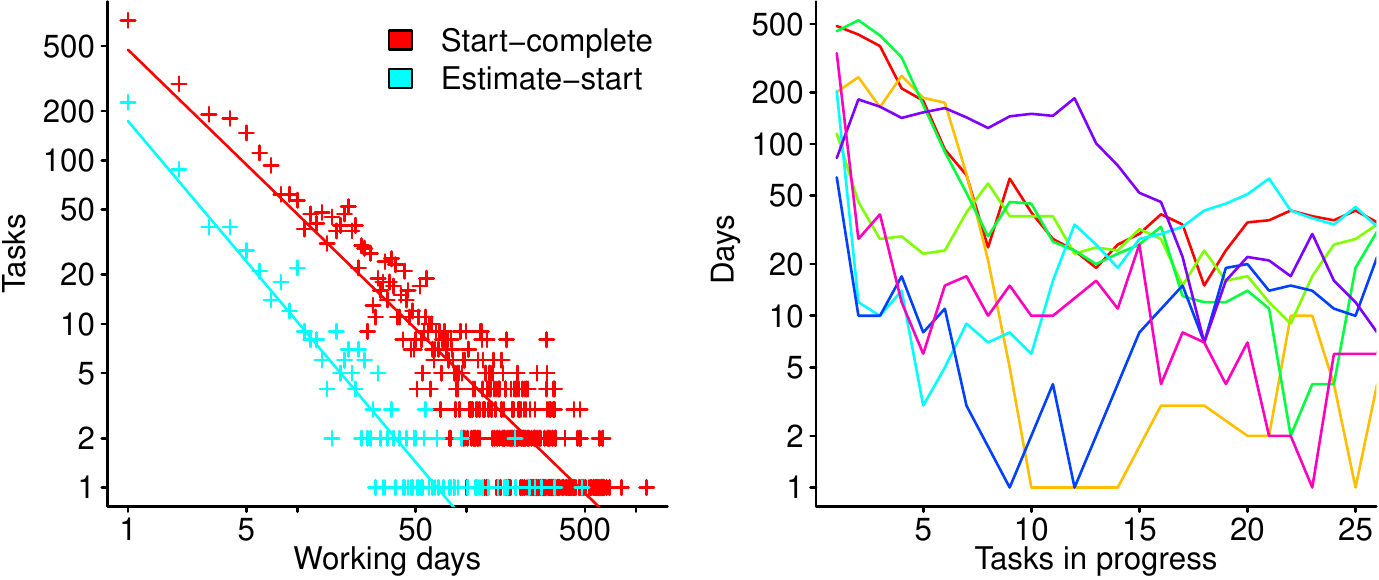}
\end{center}
\caption{Number of tasks having a given interval, in working days, between estimating/starting (blue) and starting/completing (red), with fitted regression lines (left); number of days on which a developer has started, but not yet completed, a given number of tasks; only developers who have worked on at least 200 tasks, and excluding tasks started/completed on the same day (right).}
\label{Elapsed-day:fig}
\end{figure}

The left plot in Figure~\ref{Elapsed-day:fig} shows the elapsed time,
in working days, between a task estimate being made and work starting
on the task (blue) and between work starting and completing for a
task.  The fitted regression models are (the Estimate/Start
distribution has a long tail, and the fitted model is based on
intervals of less than 100-days):

$$\mathit{Tasks}=\frac{473}{\mathit{StartCompleteWD}}$$

where: $\mathit{Tasks}$ is the number of tasks, and
$\mathit{StartCompleteWD}$ the number of working days between
estimating and starting; and:

$$\mathit{Tasks}=\frac{173}{\mathit{EstimateStartWD}^{1.2}}$$

where: $\mathit{EstimateStartWD}$ is the number of working days
between starting and completing the task.

\section{Developers and tasks}

\DJ

How many tasks do developers have 'in-progress' on any given day
(i.e., started, but not yet completed)?

The right plot in Figure~\ref{Elapsed-day:fig} shows the number of
days on which a developer (who has worked on at least 200 tasks; one
line per developer) has a given number of tasks 'in-progress'.  Tasks
started/completed on the same day are not included in the total.

\SC

Our preference was for picking a Task and working on it until
completion.  However, in an agile environment, after a cycle of work
(in our case the release), the client could (and sometimes would)
re-specify priorities.  In some cases, developing functionality could
cater for 80\% of a business problem.  It could well be decided that
this caters for a large component of the users work, so a partial
release could be arranged, and the Task is put on hold while resource
is used to address a more (perhaps new) pressing requirement.
Clarity had no way of tracking this type of behaviour.

\subsection{Corrupted tasks}

\SC

I have noticed some interesting Tasks, especially for Developer 24.
His workload consisted of a lot of repeatable work, and it looks like
he defined certain Tasks for this and just left them open!  We
automated checks for ensuring staff hours totalled 7 per working day
but there was nothing in place to check that the Tasks were being
marked as \texttt{FINISHED}.  In hindsight, this is something we
should have put in.  For the most part he treated these as normal
Tasks but just never marked them as done.  However, there are a few
catch-all Tasks whereby the Task was defined as something common
e.g., Check Log For Errors, an Estimate was set, 7 hours say BUT all
effort was flagged against that Tasks going forward.  An example is
Task \# 12640 where the estimate is 14-hours, but the actuals are
112.64, so 98.64 hours late!  He just never bothered to close the
Task and open a new one (I believe we were all guilty of this type of
behaviour, occasionally, but it obviously became the de rigueur for
some!).  This obviously has a big effect at Est vs. Act level but at
the level we worked it is just one late Task.  Toward the tail end of
SiP's trading history we were mostly supporting products, so the
review process of late running Tasks was lax as the clients did not
care as the product worked and as long as there were no surprises on
the billing, Task review was not top of their list.

\section{Monthly totals}

\DJ

\begin{figure}
\begin{center}
\includegraphics{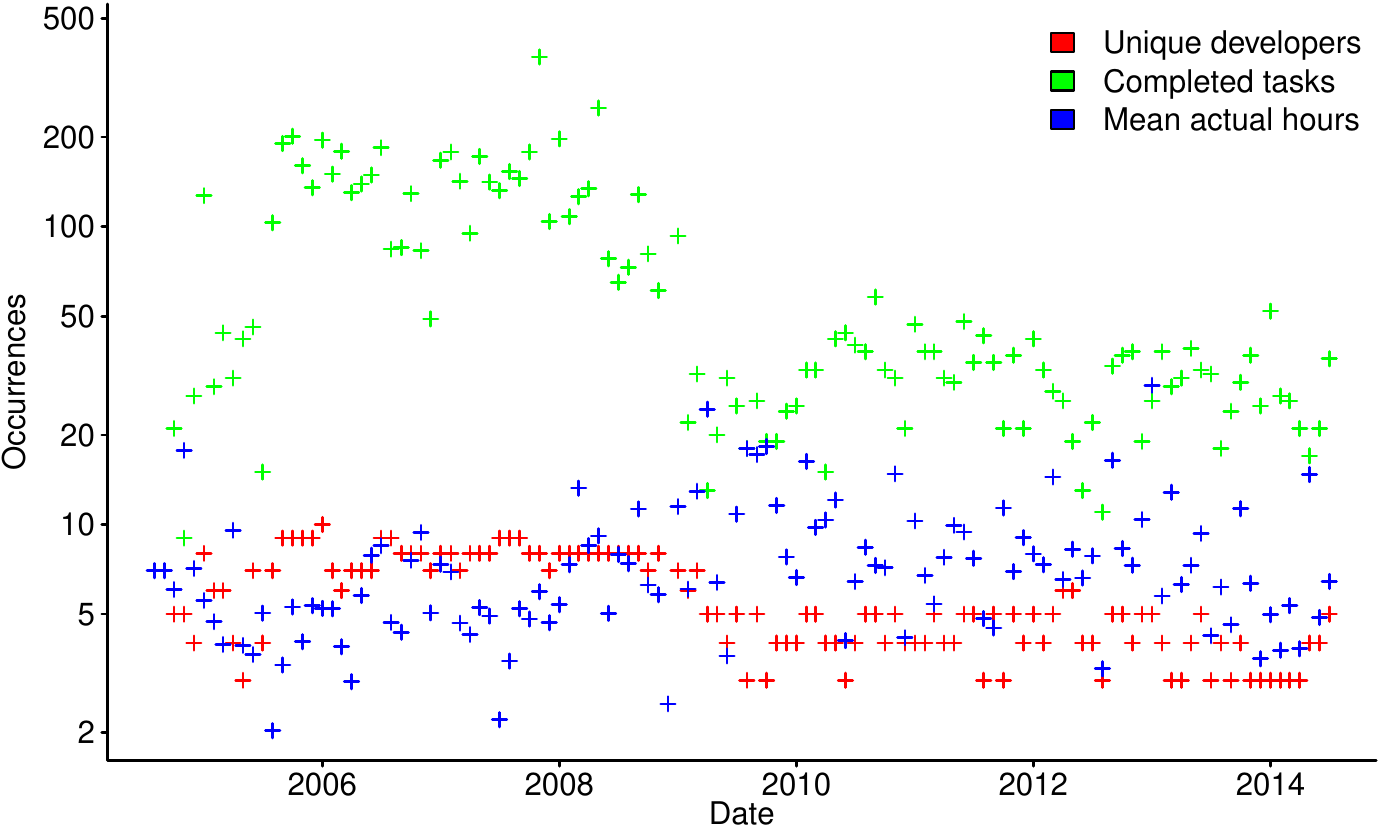}
\end{center}
\caption{Number of tasks completed per month, number of unique developers working on these tasks and average actual hours.}
\label{monthly-tot:fig}
\end{figure}

Figure~\ref{monthly-tot:fig} shows the total number of tasks
completed per month, along with the number of unique developers who
worked on at least one of these tasks.  A change-point analysis
produces three dates for the step-change in occurrences of the three
quantities plotted, and the first day of 2009 was chosen as a common
branch point.

The step-wise change around the start of 2009 raises the question of
whether there is a step-wise change in the estimate/actual models
fitted in section~\ref{first-regression-model}, which used data for
all the years; might the coefficients be significantly different for
the data pre/post 2009?

The following is the equation from a fitted model, which includes a
before/after 2009 variable:

$$\mathit{HoursActual} = \begin{cases}1.04\times\mathit{HoursEstimate}^{0.88}&\text{Before 2009-01-01}\\
		1.1\times\mathit{HoursEstimate}^{0.86}&\text{After 2009-01-01}\end{cases}$$

\section{Model fitting algorithm}

\DJ

Stephen sent me plots based on counts of the number of over/under
estimates made by specific developers.  These made me realise that
fitting a model based on equalizing the number of over/under
estimates (rather than the quantity of over/under estimate) might
provide some insights.

The standard regression modeling technique fits an equation, which can
be used to calculate the expected mean value of the response variable
(e.g., \texttt{HoursActual}).  If we are interested in calculating
the estimated median value (i.e., the estimate having an equal number
of recorded tasks above and below), quantile regression (sometimes
called median regression, although the technique also supports model
fitting based on quantiles other than 50\%) is the technique to use.

The following are the equations fitted to 75\%, 50\% (i.e., median)
and 25\% quantiles (that is, with 75\% of actual hours below the
estimate, 50\% below and 25\% below):

$$\mathit{HoursActual} = \begin{cases}1.1\times\mathit{HoursEstimate}^{1.1}&\text{75\%}\\
		1.0\times\mathit{HoursEstimate}^{1.0}&\text{50\%}\\
		0.8\times\mathit{HoursEstimate}^{0.84}&\text{25\%}\end{cases}$$

\begin{figure}
\begin{center}
\includegraphics[width=0.4\textwidth]{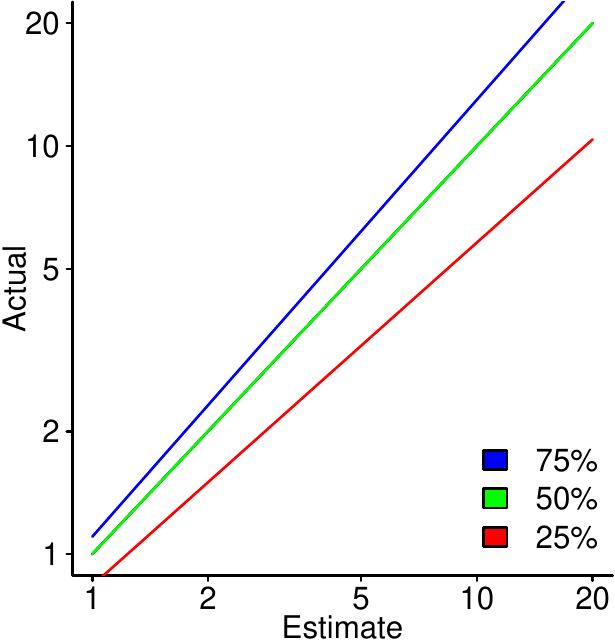}
\end{center}
\caption{Lines are from 25\%, 50\% and 75\% of a fitted quantile regression; the black line (showing Estimate==Actual) is hidden by the green line.}
\label{quant-25-50:fig}
\end{figure}

Figure~\ref{quant-25-50:fig} shows plots of these equations.  The
SipP estimates are on the 50\% (green) line, with half of all task
actuals greater/less than the estimate.  For a given task estimate,
75\% of all actuals were below the blue line (25\% below the red
line).

\section{Discussion}

\DJ

This analysis is based on 10 years of data from a company operating
within a particular niche of a particular ecosystem (financial
services).  How might companies operating today, in the same or
different ecosystems, make use of the patterns uncovered by the
analysis?

How might the findings be used?

\SC

We developed Clarity to aid our staff in providing accurate
estimates.  From a business perspective just having this information
recorded made it much easier for the client to accept our estimates.
Obviously, every client wants the work completed yesterday and can be
very inflexible when it comes to negotiating the time available to
carry out the work.  Companies can be forced into agreeing impossible
deadlines to get the work, knowing they will overrun, but it will
probably be too late for the person who appointed them to back out
without losing face (or their job), so an unholy alliance ensues,
with decreasing amity as the project overruns and corners are cut to
get it back on (impossible) track.  We found that by having evidence
of previous similar work, we could argue for more realistic timelines
from the beginning.  When we delivered (on time) this strengthened
the overall client relationship, leading to more work, rather than
SiP burning its bridges and subsequently having to find new forms of
income, which is (extremely) expensive.

Your analysis confirming that this data meets sanity checks and that
the SiP staff making the estimates are not fluctuating (too) wildly
would be used to confirm to the client that our approach through
Clarity had merit, and the data was not simply 'made up' to influence
them in directions favourable to SiP.  Essentially we could provide
evidence to potential clients that our estimates were trustworthy.

\DJ 

How accurate do estimates need to be?

\SC

That's a really good question Derek.  I think it depends on the
client and the amount of goodwill you have built up.  Some clients
are micro managers and any form of overrun is agonised over, with
lots of finger pointing and drama, which tends to taint the project
and create an us-and-them culture, which is not at all beneficial.
The best clients take a bigger picture approach, accepting that some
tasks will overrun and some will under-run.  If the aggregate is a
small under/overrun, then they generally deem the project successful.
Having a little wiggle room in each task provides protection against
those work items in which an unknown problem is being addressed, and
the likelihood of something unexpected happening is high.  Having
sufficient analysis of the problem at hand (which in turn creates a
high number of tasks) negates the need for having unnecessarily
precise estimates as things should balance out at the end of a work
cycle.

Personally, I think excessively accurate estimates (where you spend a
lot of time breaking things down to the nth degree are a waste of
time); I prefer good estimates, not guesswork (which is why we built
Clarity) but a more evidence-based approach where you can reference
aspects of work you have done before.  If you cannot find any
previous evidence for work you are currently bidding for, then I
would tread carefully as you are in unknown territory.  However,
aiming for accurate 'good' estimates should be a key goal, as it
lowers risk and means you don't have to rely on a fudge factor (i.e.,
the number of under-runs beats the number of overruns) to be able to
successfully deliver the agreed workload.  In the Lloyd's of London
marketplace 6-month overruns (seriously) are not uncommon, all
because their supplier agreed to absolutely anything to get the work.
In these instances, the client loses first and further down the line
the supplier does as their reputation is shot, and the client moves
on as the relationship is unsalvageable.

\subsection{What did I learn?}

\DJ

Many of the patterns of behavior revealed by this analysis will be
familiar to those involved in managing software development projects.
People will differ in their opinion of the impact these behaviors
have on the final outcome (i.e., the estimate); data analysis
provides an impartial view.

\begin{itemize}
\item people differ in their willingness to accept risk, and their
confidence in their own abilities.  A consequence of this difference
is making higher/lower, than average, estimates.

Developers and their managers are often aware that personal
characteristics have an impact on the estimates made.

An analysis can provide information about individual patterns of
behavior, relative to group average.  Might this information provide
useful feedback for individuals and managers?  While individual
measurement-based feedback has been provided in other work domains,
the possibility of having and using this kind of information is new
in software engineering.

\item projects have characteristics that can make it harder/easier to
accurate estimate task implementation time.

\item round numbers play a significant role in estimation.  Despite
regularly using round numbers in my own estimates, I had not
appreciated how pervasive they were; Figure~\ref{Round-Est:fig} was
eye-opening.

The habituated use of round numbers may hinder estimate accuracy,
but is it worthwhile training developers to overcome this habit?

Estimates have to be sold to clients, and if the client thinks in
round numbers a non-round number estimate introduces friction to the
sales process\cite{Loschelder_16}.  Stephen highlights the use of
project data analysis to show clients that the estimation process is
reliable, but adding a source of friction to client interactions is
never desirable.

\item developers did not learn to make more accurate estimates.

I was expecting to find evidence of learning.  It is possible
learning did take place, developers learned to increase their
productivity (with estimating accuracy remaining unchanged).  In a
competitive environment, increased productivity is a much more
desirable goal, than improved estimation accuracy.  In a
non-competitive environment, where profitability depends on hours
worked, estimation accuracy may be the learning goal, not increased
productivity.

\end{itemize}

\SC

It has been a strange experience reviewing our Clarity Task data, my
initial reaction was to try to defend any inaccurate estimates, with
the view that perhaps we were not professional enough.  On review
though, the estimates were appropriate for the environment we were
working in.  There are some hideous underestimates but at no point
did we miss a release deadline (although some functionality may have
been light or delivered as an emergency release).  The way we
normally dealt with an underestimate was those responsible just
worked overtime to commit the work (the users was therefore not
directly impacted by the slippage, we may have behaved very
differently if they were).

SiP continually developed the Lloyd's of London back office system
for a five-year period and continued to support and extend it for
another six years after that until the technical assets of the
company were sold off.  The majority of products developed using
Clarity are still in use today.

\begin{itemize}

\item Buffers to accurate estimates.

I learnt that our agile process provided a buffer to inaccurate
estimates.  By having a set number of deliverables agreed for a
release, an organisation gains the ability to balance over and under
estimates.  As long as everything promised is delivered by the
deadline then the client really does not care that some estimates
were off, you delivered what you promised, so all is good.
Intelligent release planning (not over promising and breaking the
work into enough Tasks) provides a measure of risk mitigation for
ambitious estimating.  An over running Task can also be recovered by
the developer working on it outside of normal working hours (late
nights/weekend working), which ensures it still meets the client
deadline.  Dependent on the relationship, this may be chargeable or
the supplier may have to absorb the overrun.

\item Development estimates vs. support estimates.

A more surprising finding was how SiP provided (and the client more
readily accepted) overestimates (c.f. accurate estimates) when we
moved into the support phase for the products we had built
(Figure~\ref{cou-stephen:fig} - upper plot, after 2008).  Thinking
about this, I believe one of the reasons this happened was the client
was heavily involved in the first five years of the project and would
actively review/discuss/direct work. A lot was on the line if the
project failed.  However, once the product was proven (all required
business processes were delivered) and all outsourced solutions were
successfully migrated, the project was deemed a success and the
client became more focused on other challenges. My take away here is
active client involvement (read, criticism) drives estimation
accuracy. Over estimating in the support years provided two benefits
for SiP, 1) an unexpected delay did not necessarily mean we delivered
late, 2) by passing any gains straight back to the client, we
converted cost/time savings into good will.

\item The tooling can always be better.

Our Clarity tooling has been found wanting.  It was sufficient in
ensuring that estimates and actuals were recorded but processes to
ensure that Tasks were correctly closed off, in hindsight, should
have been put into place.  Also, allowing each developer to
contribute their own estimate, rather than recording a single group
consensus would have provided better learning opportunities.

\item We became naturalised.

The analysis has made it clear that we estimated the way the client
wanted us to estimate.  We very clearly acclimatised to their working
patterns.  We would often be some of the first in the office, and
nearly always the last to leave, but providing estimates which meant
the user could begin reviewing the code close to home time or first
thing (when they would often be fielding their real work) made no
sense to them.  Providing estimates which suited the users was a
natural behaviour, when working so closely with them.

\pagebreak[4]

\item It is different at the coal-face.

Estimating at the coal-face, with all the political pressures that
brings is very different from reviewing those same estimates years
later.  We were very lucky (in the most part) with the companies and
individuals we provided solutions to.  There are some clients who
will demand the tightest of margins and then complain bitterly when
something slips, despite being warned that there was a high
probability that could happen.  We learnt over time that it was
better to decline work in certain instances rather than engage in
projects that were destined to disappoint (both parties).  Improving
estimates can however be very difficult, even with empirical evidence
if the client simply refuses to listen and insists on an unrealistic
schedule.  If you refuse the work, some other company will inevitably
pick it up.  It can be problematic explaining to the entity
commissioning the work that their expectations are unlikely to be met.

\end{itemize}

\section{Actionable point?}

\DJ

Most companies fail; Software in Partnership was a success.  A basic
tenet of engineering is: don't fix something that is not broken.

The patterns found in the data may suggest tweaks to existing
practice, and if existing patterns changes, it may indicate that
something about existing practice has changed (which may an
improvement, a regression, or neutral).

The primary actionable item for other companies, is to collect data
about what they spend their time doing.  Companies want to control
the processes they use, which is only possible when they understand
what is going on.  Patterns of behavior discovered by the analysis of
historical data can help refine existing insights or suggest new ones.

\bibliography{sipl-main}

\begin{thebibliography}{10}

\bibitem{Flyvbjerg_13}
B.~Flyvbjerg.
\newblock How planners deal with uncomfortable knowledge: {The} dubious ethics
  of the {American Planning Association}.
\newblock {\em Cities}, 32:157--163, June 2013.

\bibitem{Hatton_07}
L.~Hatton.
\newblock How accurately do engineers predict software maintenance tasks?
\newblock {\em Computer}, 40(2):64--69, Feb. 2007.

\bibitem{Heathcote_00}
A.~Heathcote, S.~Brown, and D.~J.~K. Mewhort.
\newblock The power law repealed: {The} case for an exponential law of
  practice.
\newblock {\em Psychonomic Bulletin \& Review}, 7(2):185--207, Apr. 2000.

\bibitem{Jansen_01}
C.~J.~M. Jansen and M.~M.~W. Pollmann.
\newblock On round numbers: {Pragmatic} aspects of numerical expressions.
\newblock {\em Journal of Quantitative Linguistics}, 8(3):187--201, 2001.

\bibitem{Jones_18}
D.~M. Jones.
\newblock Code \& data used in: {Evidence-based} software engineering: {Based}
  on the publicly available data.
\newblock \url{http://www.github.com/Derek-Jones/ESEUR}, 2019.

\bibitem{Jorgensen_12}
M.~J{\o}rgensen and S.~Grimstad.
\newblock Software development estimation biases: {The} role of
  interdependence.
\newblock {\em {IEEE} Trans\-actions on Soft\-ware Engin\-eering},
  38(3):677--693, May 2012.

\bibitem{Lieder_18}
F.~Lieder, T.~L. Griffiths, Q.~J.~M. Huys, and N.~D. Goodman.
\newblock The anchoring bias reflects rational use of cognitive resources.
\newblock {\em Psychonomic Bulletin \& Review}, 25(1):322--349, Feb. 2018.

\bibitem{Loschelder_16}
D.~D. Loschelder, M.~Friese, M.~Schaerer, and A.~D. Galinsky.
\newblock The too-much-precision effect: {When} and why precise anchors
  backfire with experts.
\newblock {\em Psychological Science}, 27(12):1573--1587, Oct. 2016.

\bibitem{Pope_11}
D.~Pope and U.~Simonsohn.
\newblock Round numbers as goals: {Evidence} from baseball, {SAT} takers, and
  the lab.
\newblock {\em Psychological Science}, 22(1):71--79, Jan. 2011.

\bibitem{Sonnemans_06}
J.~Sonnemans.
\newblock Price clustering and natural resistance points in the {Dutch} stock
  market: {A} natural experiment.
\newblock {\em European Economic Review}, 50(8):1937--1950, Nov. 2006.

\end{thebibliography}

\end{document}